\documentclass{jetpl}
\twocolumn

\lat

\title{Dense Quark Matter Conductivity  in  Ultra-Intense Magnetic Field}

\rtitle{Dense Quark Matter Conductivity  in  Ultra-Intense Magnetic Field}

\sodtitle{Dense Quark Matter Conductivity  in  Ultra-Intense Magnetic Field}

\author{B.\,O.\,Kerbikov$^{\dagger*}$\/\thanks[1]{e-mail: borisk@itep.ru},
 M.\,A.\,Andreichikov$^{\dagger*}$\/\thanks[2]{e-mail: andreichicov@mail.ru}}

\rauthor{Kerbikov\,B.\,O., Andreichikov\,M.\,A.}

\sodauthor{Kerbikov, Andreichikov}

\address{$^\dagger$ State Research
Center\\Institute for Theoretical and Experimental Physics,\\ 25 Bolshaya Cheremushkinskaya, 117218 Moscow, Russia\\~ \\
$^*$ Moscow Institute of Physics and Technology,\\  9 Institutskii per., Dolgoprudny, 141700 Moscow Region, Russia}

\dates{26 June 2012}{*}

\abstract{
Heavy-ion collisions generate a huge magnetic field of the order of
$10^{18} G $  for  the  duration of about 0.2 fm$/$c. This time may become an
order of magnitude  longer  if the electrical conductivity of quark matter
is large. We calculate the conductivity in the regime of high density and show
that contrary to naive expectations it only weakly depends on the magnetic field strength.}

\begin{document}

\maketitle

A large body of experimental and theoretical research conducted for over more
than ten years of operation of RHIC and the first runs of heavy ion program at
LHC has led to a revolutionary change in our view on the nature and properties
of the produced strongly interacting matter. A powerful way to investigate the
nature of a certain substance is to study its response to the external
perturbations. Few years ago it has been comprehended that in heavy ion collisions we have such
a tool. The produced quark-gluon plasma is subject to a super-strong magnetic field (MF)
generated by colliding ions which have large electric charges and are moving at
speed close to the speed of light. At the collision moment and shortly after it
$(\tau \leq 0.2$ fm$/$c) the MF reaches the value $|eB|\geq m^2_\pi \sim 10^{18} G
$ , i.e. it is of a typical QCD scale \cite{1}. In presence of quark matter
(QM) the life time of the MF may be several times longer provided the
electrical conductivity (EC) of QM is large enough \cite{2}. The properties of
QM including its transport coefficients depend on the location of the system in
the QCD phase diagram, i.e., on the value of the temperature and chemical
potential. At zero chemical potential and high temperature the EC has been
calculated by lattice Monte Carlo method yielding significantly different
results \cite{3}.Here we consider a reverse regime of high density and low or
moderate temperature. Qualitatively these conditions will be specified below.
Physically such situation may be realized in neutron stars or  in future
experiments at NICA and FAIR. On the theoretical side use can be made of the
ideas  and methods developed in condensed matter physics \cite{4}.

For a wide class of systems the EC  can be decomposed into  two contributions.
The  first  one is   Boltzmann, or Drude, and it corresponds to semi-classical
approximation when the mean free path of a particle is much larger than any
other microscopical length scale and the probabilities of a particle
interactions are added. Quantum effects enter into this contribution only via
Fermi or Bose distribution functions. A variety of quantum effects, such as the
interference of trajectories, quantum phase transitions, quantized vortices,
fluctuating Cooper pairs, Landau levels in magnetic field, etc., define the
quantum contribution to the EC. In solid state physics this contribution is
called ``quantum correction'' \cite{4} since it is inversely proportional to
$(k_Fl)$, where $k_F$ is the Fermi momentum, $l$ is the mean free path, and
$(k_Fl)\gg 1$ in solids. This is not the case in dense quark matter where
$(k_Fl)$ may even approach unity (we remind  that according to Ioffe-Regel
criterion at $k_Fl=1$ transition to Anderson localization phase occurs). The
description of the  EC in terms of the above two contributions is legitimate
for highly  disordered systems \cite{4} in the vicinity of superconducting
phase transition \cite{5}, and in AdS/CFT correspondence \cite{6}.

Our starting point  is the general expression for the EC in terms of Matsubara
Green's functions (GF) \cite{4,5}. The EC momentum and frequency dependent tensor $\sigma_{lm}(\mathbf{q},\omega_k)$, $\omega_k = 2k\pi T$ is given by:
\begin{multline}\sigma_{lm} (\mathbf{q}, \omega_k) =\\ 
\frac{e^2T}{\omega_k} \sum_{\varepsilon_n} \int \frac{\mathrm{d}^3 \mathbf{p}}{(2\pi)^3} tr \langle
  G(\mathbf{p}, \tilde{\varepsilon}_n) \gamma_l G (\mathbf{p}+\mathbf{q},\tilde{\varepsilon}_n
+\omega_k) \gamma_m \rangle.
\label{1}
\end{multline}

For $N_c=3$, $N_f=2$ we have  $e^2=3\cdot 4 \pi \alpha \left(\left(\frac{2}{3}\right)^2
+ \left(\frac{1}{3}\right)^2 \right) =0.15$. The symbol
 $\langle ...\rangle$ implies the averaging over   the disorder. For Drude EC this
 procedure is performed independently for each GF and reduces to the
 substitution of the standard Matsubara frequency $\varepsilon_n =\pi T (2n+1)$
 by
$\tilde \varepsilon_n = \varepsilon_n + \frac{1}{2\tau} sgn(\varepsilon_n)$, where
$\tau$ is the momentum relaxation time \cite{4,5}. For  the  quantum
contribution the averaging gives rise to an infinite series of the ``fan''
diagrams \cite{4,5,7} yielding the result presented at the end of this paper.
The GF has a standard form \cite{8} 
\begin{equation}
 G(\mathbf{p}, \tilde \varepsilon_n) = \frac{1}{
\gamma_0 (\tilde \varepsilon_n +\mu) - \boldsymbol{\gamma}\mathbf{p} -m},
\label{2} 
\end{equation}
where $\mu$ is the quark chemical potential. In the regime of high density and
moderate temperature the transport coefficients are dominated by the processes
occurring in the vicinity of the Fermi surface. Hence the momentum integration
in \cite{1}) is performed in the following way 
\begin{equation}
 \frac{\mathrm{d}^3\mathbf{p}}{(2\pi)^3}
=\frac12\nu \frac{d\Omega}{4\pi} d\xi,\label{3}
\end{equation}
\begin{equation}
 \nu\simeq \frac{\mu
p_F}{\pi^2} \left[ 1+\frac13 \left(\frac{\pi
T}{\mu}\right)^2\right],
\label{4}
\end{equation}
where $\xi=(\mathbf{p}^2+ m^2)^{1/2} -\mu$.    Calculation of the Drude EC reduces to the evaluation  of  a one--loop diagram
defined by (\ref{1}--\ref{4}). First we perform the $tr$ operation over Dirac
induces, then integration in the complex $\xi$--plane, and finally the
$\varepsilon_n$ summation. Replacing the Matsubara frequency  $\omega_k$ by the
physical ones $\omega_k=-i\omega$, we write down the resulting expression
for the frequency dependent Drude EC 
\begin{equation}
\sigma_{ll} (\omega) =\frac23 e^2\nu
v_F \left( \frac{\tau}{1+\omega\tau}\right),\label{5}
\end{equation}
 where $v_F=p_F/\mu$.
The antiquark contribution  is damped near the Fermi surface and dropped in Eq.
(\ref{5}). If in (\ref{5}) we replace $\mu \to \tilde \mu + m$, and take the
limit $|\mathbf{p} | \ll m$, we arrive  to the standard Drude formula with $\nu =
mp_F/\pi^2$ and $\tilde \mu$ being the non-relativistic chemical potential. For
orientation purposes let us  estimate $\sigma(\omega \to 0)$ for the following
set of parameters: $\mu=400 $ MeV, $T=100$ MeV, $v_F=1, \tau=0.8$  fm. One gets
$\sigma\simeq 0.04$ fm$^{-1}$. To our knowledge, the EC has not been calculated
before in this domain of the QCD phase diagram. Results obtained at $\mu = 0$
and  different values of $T$ \cite{3} differ from each other by an order of
magnitude. Our value  $\sigma\simeq 0.04$ fm$^{-1}$ lies within the interval of
the EC values given in \cite{3}.

An important point is that Eq. (\ref{5}) contains a large parameter $\mu\sim
300-500$ MeV. As  we shall see this leads to the stabilization of the  EC in MF up
to $(eB/\mu)\tau\sim 1$. Let the constant MF $\mathbf{B}$ be directed along the
$z$-axis. The Drude EC can be evaluated either from the one-loop diagram with
MF entering into the propagators \cite{9}, or from the Boltzmann kinetic
equation. Diagrammatic calculation is more cumbersome, both methods lead to the
same result which is an anticipated generalization of (\ref{5}). We present the
result leaving the straightforward but lengthy derivation for the forthcoming
detailed paper. In line with the symmetry requirements the EC along $z$--axis
remains unchanged while the transverse one is equal to 
\begin{equation}
 \sigma_\bot(\omega
= 0, \Omega) = \frac{\sigma_0}{1+\Omega^2\tau^2},\label{6}
\end{equation}
 where $\sigma_0\equiv\sigma_{ll}(\omega=0)$ given by (\ref{5}), and $\Omega=eB/\mu$.
For the set of parameters considered above, namely for $\mu=400$ MeV,
$\tau=0.8$ fm we have $\Omega^2\tau^2 <1$ up to $eB<5 m^2_\pi$.

Now we give  a cursory glance on the quantum part $\sigma'$  of the EC. As
already mentioned,  it includes the  interplay of various quantum phenomena. In
the regime  under consideration  the major role is played by the formation of
the fluctuation (or precursor) Cooper pairs \cite{5,10}. Again we refer to the
forthcoming  paper analyzing diagrams containing propagators of such pairs. The
order-of-magnitude estimate of quantum EC is $|\sigma'| \sim e^2/\pi^3l$, where
$l=v_F\tau$ is  the mean free path, the sign of $\sigma'$ may be either
positive, or negative. When $\sigma'<0$ and large by the absolute value, the
system approaches the Anderson localization regime. When $B\neq 0$ the above
estimate transforms into $|\sigma'| \sim e^2/\pi^3l_B$, where $l_B$ is the magnetic
length, $l_B=(eB)^{-1/2}$. Therefore $\sigma'$ is not proportional to $eB$ as
might be naively expected from the fact that the number of modes degenerating
in a unit transverse area is proportional to B.

To summarize, we have for the first time evaluated the EC of dense relativistic
quark matter with low and moderate temperature. We have demonstrated the
rigidity of the EC on the magnetic field. The details of the calculations will
be presented in the forthcoming paper.

 B.K. is indebted to A.Varlamov  for  illuminating  discussions and  to
L.Levitov  for the remarks. Support from RFBR grants 08-02-92496-NTSNIL-a and
10-02093111-NTSNIL-a are gratefully acknowledged.

\begin{center}
\emph{Remark and references added in v.3:}
\end{center}

The factor $[ 1+ \frac{1}{3}(\frac{\pi T}{\mu})^2]$ in
(\ref{4}) is not a corollary of the $\xi$-integration.It arises in more general approaches-see \cite{11,12}.The authors thank E.Megias for this remark. We also note that our conclusion on the rigidity of the
EC on the magnetic field was confirmed by a very recent calculation \cite{13}
in the instanton liquid model.

\end{document}